\documentclass[aps,pra,twocolumn,showpacs,amsmath]{revtex4}
\usepackage{graphicx}
\usepackage{amssymb}

\newcommand{\bra}[1]{\langle #1 | \,}
\newcommand{\ket}[1]{\, | #1 \rangle}

\newcommand{\be}{\begin{equation}}
\newcommand{\ee}{\end{equation}}
\newcommand{\bea}{\begin{eqnarray}}
\newcommand{\eea}{\end{eqnarray}}
\newcommand{\besa}{\begin{subequations}\begin{eqnarray}}
\newcommand{\eesa}{\end{eqnarray}\end{subequations}}

\newcommand{\sgn}{\operatorname{sgn}}

\begin{document}
\title{Lattice two-body problem with arbitrary finite range interactions.}
\author{Manuel Valiente}
\affiliation{Institute of Electronic Structure \& Laser, FORTH, 
71110 Heraklion, Crete, Greece}
%\affiliation{Institut f\"ur Physik, Humboldt-Universit\"at zu
%  Berlin, Newtonstr. 15, D-12489 Berlin, Germany}
\date{\today}

\begin{abstract}
We study the exact solution of the two-body problem on a tight-binding
one-dimensional lattice, with pairwise interaction potentials which have an arbitrary but
finite range. We show how to obtain the full spectrum, the bound and
scattering states and the ``low-energy'' solutions by very efficient and easy-to-implement numerical
means. All bound states are proven to be characterized by roots of a polynomial whose degree depends linearly on the range of the potential, and we
discuss the connections between the number of bound states and the scattering
lengths. ``Low-energy'' resonances can be located with great precission with
the methods we introduce. Further generalizations to include more exotic interactions are also discussed.   
\end{abstract}


\maketitle

%%%%%%%%%%%%%%%%%%INTRODUCTION%%%%%%%%%%%%%%%%%%%%%%%
\section{Introduction}

The study of quantum mechanics in periodic structures is one of the central
topics in condensed matter physics since many decades \cite{Ashcroft}. The
behavior of electrons in a crystal or, more generally, interacting particles
in a periodic potential, even at the few-body level, puts forward a major
theoretical and numerical challenge for theorists. Therefore, oversimplified
models which are still able to capture certain qualitative features of the
original problem have been proposed. The so-called Hubbard model
\cite{Hubbard} for electrons in a metal and its bosonic counterpart
\cite{bosMI,Jaksch}, assume that the particles only populate a single energy band of
the periodic potential and that the effective interaction has a short range
(if not zero range) character. 
Recent advances in the physics of ultracold atoms in optical lattices
\cite{OptLatRev} have opened a fascinating framework in which it is possible
to simulate with unprecedent accuracy some of the models traditionally used in
condensed matter physics. In particular, the transition from a superfluid to a Mott
insulator of bosons loaded in an optical lattice has been successfully
observed \cite{transition}. In another spectacular experiment \cite{Winkler}
, repulsively
bound pairs of atoms have been produced, and their main properties have been
measured. This experiment has stimulated renewed interest in the study of
few-body effects in discrete lattices
\cite{PiilMolmer,MVDP1,MVDP2,JinSong,weiss,MVDPAS,PiilMolmerdifferentmass}, which had
been almost only 
studied by Mattis {\cite{RevMattis}} about twenty years ago.

Both two-body problems on a one-dimensional tight-binding lattice with on-site
\cite{Winkler,MVDP1,PiilMolmer} as well as nearest-neighbor \cite{MVDP2}
interactions can be solved exactly. Several conclusions about them can be obtained \cite{MVDP2}: (i) Bound states can be calculated via a certain
polynomial equation of different degree depending on the range of the
interaction. (ii) The scattering states, both symmetric (bosonic) or
antisymmetric (spin-polarized fermionic) are well described, asymptotically,
by a single phase shift which depends again on the range of the
potential. (iii) The ``low-energy'' properties of those systems, characterized
by the scattering lengths, can always be calculated and appear to be rather
simple expressions of the respective interaction potentials.

Therefore, the next relevant question to ask is if there is a general pattern
followed by the solutions and properties of the two-body problem when the
interactions are of arbitrary but finite range. In this article we deal with
this question and find that, indeed, there is a well defined pattern for the
bound states, and that all scattering properties can be calculated very
efficiently. For the particular, yet very important case of ``low-energy''
scattering 
we show how to obtain the scattering lengths very accurately even
without knowing the phase shift in general. We treat both 
identical and distinguishable particles which can, of course, have different
tunneling rates, thanks to a generalization of the center of mass separation
ansatz for the two-body problem. To illustrate our results, we apply them to a
model dipolar potential with a cutoff at a certain, long enough range.

%%%%%%%%%%%%%%%%%%%%%%%%%%%%%%SEPARATION%%%%%%%%%%%%%%%%%%%%%%%%%%%%%%%% 
\section{General separation of the two-body problem}

We consider two particles, labeled A and B, in general having different tunneling
rates \cite{footnote1} $J_A$ and $J_B$, and interacting via a symmetric two-body
potential $V(z)=V(-z)$. The reduction of the two-body to a one-body problem
was first carried out in \cite{PiilMolmerdifferentmass,Martikainen}.

The two-body one dimensional discrete Schr\"odinger
operator $H$ for the two-body system, acting on a two-body wave function $u$ in
$\ell^2(\mathbb{Z})\otimes
\ell^2(\mathbb{Z})$ is, in first-quantized form,
\begin{align}
(Hu)(n_A,n_B)&= \nonumber\\
              &-J_A\left[u(n_A+1,n_B) + u(n_A-1,n_B)\right]\nonumber\\
              &- J_B\left[u(n_A,n_B+1) + u(n_A,n_B-1)\right] \nonumber\\
              &+ V(|n_A-n_B|)u(n_A,n_B),
\end{align}
where $n_A$ and $n_B$ are the (integer) lattice positions of particle $A$ and
$B$, respectively. For simplicity, we set the lattice spacing $s\equiv 1$, so
that distances, lengths and quasi-momenta are dimensionless.
For the moment, we do not allow $V$ to become infinitely large, that is, $|V(|n|)|<\infty$ for all $n\in
\mathbb{Z}$; this condition can be relaxed by allowing $|V(0)|\to \infty$ and then
applying the Bose-Fermi mapping theorem (BFMT) \cite{Girardeau} once the
problem is reduced to a one-body equation. Moreover, we assume that $V$ is an arbitrary finite
range potential of range $\rho\in \mathbb{Z}$, that is, $V(|n|>\rho)=0$ with at least $V(\rho)\neq
0$.\\
In order to solve this problem exactly we need to transform the Hamiltonian $H$ to a
single particle operator. For this purpose, consider the ansatz
\be
u(n_A,n_B)=u_K(z) e^{-i\beta_K z+iKR},\label{generalseparation}
\ee
 where $R=(n_A+n_B)/2$ and $z=n_A-n_B$ are, respectively, the center of mass
 and relative coordinates; $K$ is the total quasi-momentum and
\be
\tan{\beta_K}=\frac{J_A-J_B}{J_A+J_B}\tan{(K/2)}.\label{tanbetaK}
\ee
Note that when $J_A=J_B\equiv J$, the ansatz
(\ref{generalseparation}) reduces to the well known case of identical
particles \cite{Winkler,PiilMolmer,MVDP1}. 
By inserting the choice (\ref{generalseparation}) in the Schr\"odinger
equation $Hu=Eu$ we arrive at the desired
single-particle Hamiltonian for each value of the total quasi-momentum $K$
\be
(\tilde{H}u_K)(z)=-|J^{(K)}|[u_K(z+1)+u_K(z-1)]+V(|z|)u_K(z),\label{ham}
\ee
where the so-called collective tunneling rate \cite{PiilMolmerdifferentmass} has the form
\be
|J^{(K)}|=\sqrt{J_A^2+J_B^2+2J_AJ_B\cos{K}}.
\ee
Note that the reduced Hamiltonian of Eq. (\ref{ham}) is equivalent to the
findings in \cite{PiilMolmerdifferentmass,Martikainen}. 
At this point it is convenient to introduce an adimensional Hamiltonian by dividing
it by the collective tunneling, which is equivalent to setting
$J^{(K)}\equiv 1$ (energies become dimensionless) in Eq. (\ref{ham}), and rename $u_K\equiv u$ for
simplicity. We will assume this in the subsequent discussions.

%%%%%%%%%%%%%%%%%%%%%%%%%%%%%BOUND STATES%%%%%%%%%%%%%%%%%%%%%%%%%%%%%%%%%%%%%%%%%%%%%%%%%%%%
\section{Bound states}\label{boundgeneral}

We pursue the exact solution for the bound states of any two-body system on
the lattice with finite range interactions. Before doing so, we need to define
what is actually meant by bound state, mathematically, for the convenience of
the reader.

We define a bound state of $\tilde{H}$, Eq. (\ref{ham}), as any square-summable solution $u(z)$ of the
discrete time-independent Schr\"odinger equation $\tilde{H}u=Eu$ with its associated
eigenvalue $E$ lying
outside the essential spectrum of $\tilde{H}$,
$\sigma_{\mathrm{ess}}=[-2,2]$. Recall that ``outside the essential spectrum'' can
actually mean above \cite{Winkler,PSAF,PiilMolmer,MVDP1} and not only below
the continuum. 

It is
already known that for any finite range potential $V$ there exists at least
one symmetric bound state \cite{Damanik}; it is also known that the maximum
number of symmetric (antisymmetric) bound states of $\tilde{H}$ is $\rho+1$
($\rho$) \cite{Teschl}. Now we show rigorously {\it how} to calculate all these
bound states exactly. The formulation of this result is as follows:\\
{\it Theorem.\\
Let $\tilde{H}$ be the Hamiltonian (\ref{ham}) with $V$ a range-$\rho$ ($<\infty$)
potential. Then all bound states $u(z)$ of $\tilde{H}$ have the decay property
$u(z)\propto\alpha^{|z|-\rho}$ for $|z|\ge \rho$, $0<|\alpha|<1$; the energies of the bound states are
given by $E=-\alpha-1/\alpha$. If $u(z)$ is symmetric then $\alpha$ is a root of a polynomial of
degree $2\rho+1$ if $\rho\ge 1$ and, if $\rho=0$, its degree is $2$; if $u(z)$ is
antisymmetric and $\rho>0$ then $\alpha$ is the root of a polynomial of degree
$2\rho-1$.}

{\it Proof}. Applying the exponential ansatz for $u(z)$ with $|z|\ge
\rho$ yields immediately 
\be
E=-\alpha-1/\alpha\equiv f(\alpha).
\ee
 Since
$f((-1,0)\cup (0,1)) = (-\infty,-2)\cup (2,\infty)$ and $f$ is injective in
$(-1,0)\cup(0,1)$, we have that the exponential ansatz is the only possible form
 for the bound states outside the range of $V$.\\  
To see that $\alpha$ is a root of
a polynomial one shows by induction, for $\rho\ge 2$, that if $u(z)$ is exponentially decaying,
then $\alpha^n u(\rho-n)=Q^{(n)}_{2n-1}(\alpha)$ and $\alpha^{n-1}
u(\rho-n-1)=Q^{(n-1)}_{2n-3}(\alpha)$, where $Q^{(m)}_k$ are polynomials of
degree $k$. For symmetric solutions the polynomial equation is then obtained
by setting $u(1)=u(-1)$ and, for
antisymmetric solutions, by setting $u(0)=0$, which proves our
statement. For $\rho=0$ and $\rho=1$ the result can be proved by explicitly
obtaining the polynomial equation \cite{MVDP1,MVDP2}.

The theorem presented here implies that for any finite range potential one has to solve
a polynomial equation whose degree grows slowly with increasing $\rho$. The way
of obtaining such polynomials is, as can be observed from the proof, inductive: we start by setting $u(\rho)=1$  
and proceed to calculate $u(\pm 1)$ and $u(0)$ by recurrence and solve the
respective symmetry constrains $u(1)=u(-1)$ or $u(0)=0$. Certainly if $\rho$
gets too large it becomes inconvenient to get such polynomials for a general
potential $V$, and in this case we should obtain the coefficients
of the polynomial for the given particular potential.

\begin{figure}[t]
\includegraphics[width=0.38\textwidth]{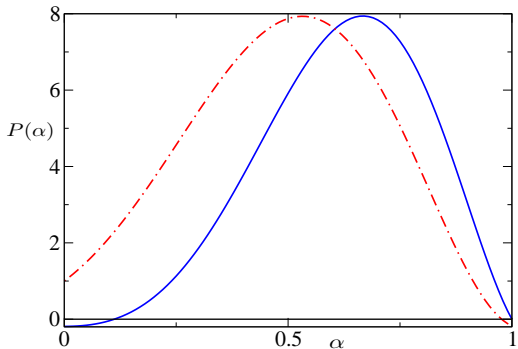}
\caption{(Color online). Polynomials for symmetric (blue solid line) and antisymmetric (red
  dashed-dotted line) bound states with the interaction (\ref{Vz3}), whose roots $\alpha$ characterize the bound states
  with energy $E=-\alpha-1/\alpha$. The horizontal line corresponds to zero ordinate.}
\label{fig:FIG2-general}
\end{figure}

Consider now the specific choice of the potential
\be
V(z)=\left\{
\begin{array}{rl}
-\frac{1}{|z|^3} & \text{if } 0<|z|\le 10 \\
0 & \text{if } |z|>10 \\
-9.7313 & \text{if } |z|=0, \label{Vz3}
\end{array} \right.
\ee
which corresponds to a dipole-dipole interaction with a cutoff at a finite but
long range, and where the divergence of the potential at $z=0$ has been
substituted by a finite value. Note that such dipolar interactions, with a
tunable on-site interaction $V(0)$, can be realized with dipolar atoms or
molecules in optical lattices \cite{Trefzger1,Trefzger2,Menotti}.
We have
calculated both polynomials $P(\alpha)$ for the symmetric and
antisymmetric bound states numerically, with their roots characterizing the
bound states. The results are shown in Fig. \ref{fig:FIG2-general}. For symmetric bound
states the polynomial has only one root in $(-1,1)$, and
therefore only one bound state \cite{footnote2}. The polynomial has a root at $\alpha=1$, which
means that it has low-energy resonance. We will discuss these resonances in Section \ref{scatlengthgeneral}. The polynomial for antisymmetric bound states has also
one and only one root in $(-1,1)$, in agreement with the discrete Bargmann's
bound \cite{HundertmarkSimon}.
%%%%%%%%%%%%%%%%%
%%%%%%%%%%%%%%%%%

%%%%%%%%%%%%%%%%%%%%%%%%%%%%%SCATTERING STATES%%%%%%%%%%%%%%%%%%%%%%%%%%%%
\section{Scattering states}

\begin{figure}[t]
\includegraphics[width=0.38\textwidth]{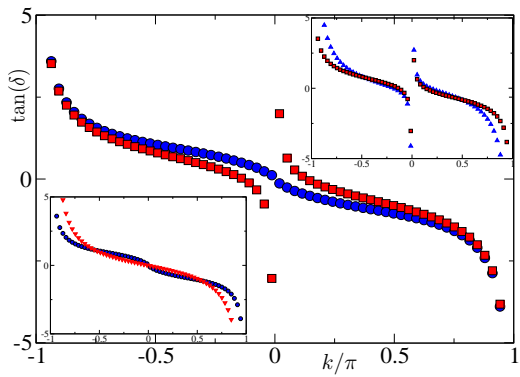}
\caption{(Color online). The calculated phase shifts
  ($\tan(\delta)$ with $\delta=\delta_S$ or $\delta=\delta_A$)
  for symmetric (blue circles) and antisymmetric (red squares) scattering
  wave functions, Eqs. (\ref{scatteringsym}) and (\ref{scatteringant}), as functions of the
relative quasi-momentum $k$, for the potential (\ref{Vz3}). Left inset:
comparison of the symmetric phase shift (blue circles) with the one obtained
with a model range-1 potential (red triangles), Eq. (\ref{modelpot}). Right
inset: antisymmetric phase shift (red squares) compared to the result with the
model range-1 potential (blue triangles). The axes of the insets have the same
meaning as those of the main figure.}
\label{fig:FIG3-general}
\end{figure}

After having introduced the first main result of this paper, which deals exclusively with
 bound states, it is natural to ask about the
exact scattering properties of the system. For finite range potentials, the scattering states of the Hamiltonian $\tilde{H}$ are
asymptotically plane waves, that is, for $|z|\ge \rho$ we have 
\begin{align}
u_S(z)&\propto \cos (k|z|+\delta_S),\label{scatteringsym}\\
u_A(z)&\propto \sgn(z) \cos(k|z|+\delta_A),\label{scatteringant}
\end{align}
where $S$ and $A$ denote, respectively, symmetric and antisymmetric
solutions. Their associated eigenenergies are given by the well known
tight-binding energy dispersion relation \cite{Ashcroft}
\be
E=-2\cos(k).
\ee
However, a general result concerning 
the phase shifts $\delta_S$ and $\delta_A$ does not seem feasible, and it is quite cumbersome to obtain them in closed
form for long enough ranges. One can, however, calculate the phase shifts (and
from them the exact solution at all $z$) numerically by recurrence. To this
end, we set $u_S(\rho+1)=\cos(k(\rho+1)+\delta_S)$ and
$u_S(\rho)=\cos(k\rho+\delta_S)$ and analogously for antisymmetric solutions,
from equations (\ref{scatteringsym}) and (\ref{scatteringant}). Then we calculate
$u_S(-1)$ and $u_S(1)$ for symmetric solutions with the help of the
Schr\"odinger equation (\ref{ham}) and solve $u_S(-1)=u_S(1)$. In the case of
antisymmetric solutions, the relevant equation is $u_A(0)=0$. We have done so
for the example potential of Eq. (\ref{Vz3}), as is plotted in
Fig. \ref{fig:FIG3-general}. There, we clearly observe that the main
differences between both phase shifts occur at low quasi-momenta where the
symmetric solution is resonant (see Fig. \ref{fig:FIG2-general}); looking at slightly higher quasi-momenta already
shows good agreement between both phase shifts. This means that far from $k=0$
fermionization appears rapidly: the large on-site interaction $V(0)$ in (\ref{Vz3}) acts as a
hard-core at high energies, for which the resonance plays no role (it is located
at the bottom of the continuum), and therefore the symmetric phase shifts are
close to the antisymmetric (``fermionic''). At low quasi-momenta, the resonance
obviously dominates the asymptotic behavior of the symmetric scattering states. 

In the insets of Fig. \ref{fig:FIG3-general}, we plot the
comparison of the phase shifts for the potential in Eq. (\ref{Vz3}) and a
model range-1 potential, whose analytic solution is known \cite{MVDP2}. The model potential $W$ is chosen so
as to be consistent with Bargmann's bound \cite{HundertmarkSimon}, and to be resonant for
the lowest-energy symmetric solution. We obtain \cite{MVDP2} 
\begin{align}
W(1)&=-\sum_{z=1}^{10}z^{-3}=-1.19753\nonumber\\
W(0)&=-12.125.\label{modelpot}
\end{align}
The qualitative agreement between the results using $V$ or $W$ for the
symmetric eigenstates is manifest in Fig. \ref{fig:FIG3-general} and, as
expected, the differences are most noticeable in the high quasi-momentum
regime. For antisymmetric eigenstates the agreement is very good, even
quantitatively, until $|k|\simeq \pi/2$. The simplified potential (\ref{modelpot})
can thus be used as a good approximation for the interaction (\ref{Vz3}) in
the problem of many (spin-polarized) fermions or
hard-core bosons on a one dimensional lattice at low energies (around the
ground state) and low filling (typically much smaller than half the number of
lattice sites). Such a problem can then be solved exactly by means of the
Bethe ansatz \cite{BetheI}.      
%%%%%%%%%%%%%%%%%

%%%%%%%%%%%%%%%%%SCATTERING LENGTHS%%%%%%%%%%%%%%%%%%%%%%%%%%%%%%%%%5
\section{Scattering lengths and zero-energy resonances}\label{scatlengthgeneral}

%%%%%%%%%%%%%%%%%%%%%%%%%%low-energy%%%%%%%%%%%%%%%%%%%%%%%%%%%%%%%%%
\subsection{Low-energy scattering}

The ``low-energy'' ($k\to 0,\pi$) scattering properties of the
two-body system can be understood via a simple, yet exact, calculation of the
scattering lengths. Indeed, the solution of the
time-independent Schr\"odinger equation when $k\to 0$ ($k\to \pi$) has an energy
$E=-2$ ($E=+2$), and has the asymptotic ($|z|\ge \rho$) behavior
\begin{align}
u_S(z) &= (\mp 1)^z \frac{|z|-a_S^{\pm}}{\rho-a_S^{\pm}}\\
u_A(z) &= \sgn(z) (\mp 1)^z \frac{|z|-a_A^{\pm}}{\rho-a_A^{\pm}} ,
\end{align}
where $a_i^{-}$ ($a_i^{+}$) is the scattering length at $k\to 0$ ($k \to
\pi$), $i=S,A$. It
must be noted that, in the case of the lattice, there are {\it four} different
scattering lengths, two for ``bosons'' (symmetric solutions) and two for
spin-polarized ``fermions'' (antisymmetric solutions). In
order to calculate the scattering lengths we proceed as follows : using
the recurrence relation from $z=\rho$ by setting $u_i(\rho+1)=(\mp 1)^{\rho+1}
[1+1/(\rho-a^{\pm})]$,
$u_i(\rho)=(\mp 1)^{\rho}$ and $E\equiv E_{\pm}=\pm 2$, the scattering lengths for the symmetric states are obtained
by solving the equation
$(V(0)-E_{\pm})u_S(0)-2u_S(1)=0$ (see proof of the theorem), while for
the antisymmetric states the
equation to solve is $u_A(0)=0$. It is remarkable that the resulting
equations for the scattering lengths as functions of the potential can be cast
as linear in $a^{\pm}$, that
is, are of the form $s_0a^{\pm}+b_0=0$ with $s_0$ and $b_0$ real constants
which depend on $V(z)$. In fact, this is an alternative way of defining the
four lattice scattering lengths, totally equivalent to the definition
$a^{\pm} = -\lim _ {k \to \pi,0} \partial_k \delta$ \cite{PiilMolmerFeshbach}, with the advantage of not
needing to know the phase shift explicitly. It must be noted
at this point that, strictly speaking, scattering lengths are unique of one-dimensional lattices
since the radial symmetry is lost in dimensions $d>1$. 

\begin{figure}[t]
\includegraphics[width=0.38\textwidth]{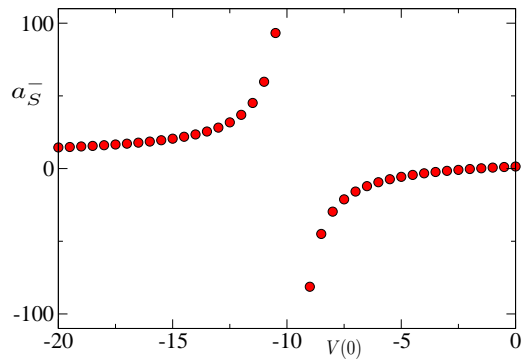}
\caption{(Color online). Scattering length $a_S^-$
  (see text) as a function of the free parameter $V(0)$, with $V(0<|z|\le
  10)=-1/|z|^3$ and $V(|z|>10)=0$.}
\label{fig:FIG1-general}
\end{figure}

As an example, we have calculated $a_S^-$ for the dipolar potential with
a cutoff as $V(0<|z|\le \rho)=-1/|z|^3$, $V(|z|>\rho)=0$, and leaving $V(0)$
as a free parameter. The results are shown in Fig. \ref{fig:FIG1-general} for
a range $\rho=10$, where there is a resonance clearly marked at the point were
the scattering length diverges.
%%%%%%%%%%%%%%%%%
%%%%%%%%%%%%%%%%%

The divergence of one of the scattering lengths can happen for different values of the total
quasi-momentum $K$ \cite{footnote3}. In the simplest case of a zero range interaction with
$V(0)\equiv U$, the system is known to have no ``zero-energy'' 
resonances \cite{MVDP1}. For longer ranges, already starting with $\rho=1$
\cite{MVDP2}, these resonances can occur. With the method outlined in this
work we are able to predict when, for a given range-$\rho$ potential with one or
more free parameters $\{V(z_1),V(z_2),\ldots,V(z_n)\}$, there is
such a resonance. To do so, one sets $u(z\ge \rho)=(\mp 1)^z$ and iterates recursively as
has been explained, and then solves the resulting equation for symmetric and
antisymmetric states, getting a relation among the free parameters of
the potential to have a resonance.\\
We consider again the example of Fig. \ref{fig:FIG1-general}. As we
have already noted, the system can admit one resonance at the bottom of the continuum for
the symmetric states. From Fig. \ref{fig:FIG1-general}, the approximate location
of the resonance can be inferred. However, once the scattering length starts
to diverge, an accurate location of the resonance grafically is very hard (if
not impossible), especially if the resonance is very sharp as a function of
$V(0)$. With our method, we are able to locate the resonance
very precisely, obtaining for the example we are dealing with a value of $V(0)
= -9.7313$ \cite{footnote4}. This is exactly the value chosen in the previous sections to
match this resonance.

%%%%%%%%%%%%%%%%%%%%%%%%%BOUND STATES CLOSE TO A RESONANCE%%%%%%
\subsection{Near-resonant bound states}
It is well known that the binding energy and size of a near-resonant bound
state (NRBS) are closely related to the (large) scattering
length $a^{\pm}\lessgtr 0$. On a one dimensional
lattice, the relation between the binding energy $E_b$ and the scattering
length in the effective mass approximation is given by 
\be
|E_b| \approx \frac{\hbar^2}{2|m^{\pm}|(a^{\pm})^2},\label{Eb}
\ee
where $m^{\pm} = \mp \hbar^2/2$ is the effective mass of the pair with an
obvious notation. Note that Eq. (\ref{Eb}) is also valid in the context of
Feshbach resonances \cite{PiilMolmerFeshbach}.

When the scattering length is large, the parameter $\alpha$ characterizing
a NRBS is close to $\pm 1$ (for $E=-2$ and $E=2$,
respectively). By writing $|\alpha|=1-|\xi|$, it is not difficult to show that
the size of the bound state, for $\rho \ge 1$, behaves as $\ell ^* \equiv \langle |z| \rangle \sim 1/(2|\xi|)$ for $|\xi| \ll 1$. Mathematically, one has
\be
\lim_{\xi \to 0} |\xi| \ell^* = \lim_{\xi \to 0} |\xi| \frac{\bra{u} |z| \ket{u}}{\Vert u \Vert ^2} = \frac{1}{2}.
\ee     
Since for small $|\xi|$, $|E_b|\approx \xi^2$, we have that the
relation between the size of the NRBS and its associated scattering length is
given by
\be
\ell ^* \approx \sqrt{\frac{|m^{\pm}|}{2\hbar^2}} |a^{\pm}| = \frac{|a^{\pm}|}{2},
\ee
showing that the size of the NRBS grows linearly with the scattering length,
as we expected.

%%%%%%%%%%%%%%%%%%%%%number of bound states%%%%%%%%%%%%%%%%%%%%5
\subsection{The number of bound states}

The scattering lengths are very useful quantities in the sense that knowing their
precise values implies knowing the total number of bound states with energies lying below or above
the continuum. For this purpose, we use the discrete analog of Sturm
oscillation theory \cite{Teschl}. In simple terms, oscillation theory states
that the number of nodes \cite{footnote5} of the zero-energy ($E=E_{-}=-2$)
symmetric (antisymmetric) solution
$u$ of $\tilde{H}u=Eu$ in $\mathbb{Z}_+$ is exactly the number of
symmetric (antisymmetric) eigenstates
below $E_{-}=-2$. Since any state with energy below $E_{-}$ is a bound state, the
number of nodes of $u$ is the number of bound states below the continuum. To see
how many bound states there are with energies above the continuum, we make use of the
transformation $\hat{G}$ of Appendix \ref{app}, Eq. (\ref{trans}), or,
equivalently, count the number of missing nodes. 

We apply oscillation theory now to our example with the potential of
Eq. (\ref{Vz3}) leaving again $V(0)$ as a free parameter. We calculate (not shown) the symmetric zero-energy solution
for a given scattering length $a_S^-$ with the methods introduced in this
section and see that for $V(0)<-9.7313$ there are two bound states with energies
below $E_{-}$ and for $V(0)\ge -9.7313$ there is exactly one bound state below the
continuum. 

%%%%%%%%%%%%%%%%%%%%%%%%%%EXCHANGE%%%%%%%%%%%%%%%%%%%%%%%%%%%%%%%%%%
\section{A generalization}

More general Hamiltonians with exchange
operators appear when dealing with the problem of one ``free'' boson and a
bound pair \cite{MVDPAS}. In such 
case, the effective particles are distinguishable, there is a hardcore on-site
interaction, an effective range-1 potential and a first order exchange term. We assume now
that we have the following general one-body Schr\"odinger operator
\begin{align}
(H_{\mathrm{ex}}u)(z)=&-[u(z+1)+u(z-1)]+V(|z|)u(z)\nonumber\\
&+\Omega(|z|)(\hat{P} u)(z),
\end{align}
where $V(0)$ can be finite or infinite and where $\hat{P}$ is the discrete parity
operator. We further assume that $\Omega(|z|)$ has a finite range $\rho_{\mathrm{ex}}$
with no on-site exchange, $\Omega(0)=0$, since it can be included in $V(0)$.

Obviously $[\hat{P},H_{\mathrm{ex}}]=0$, and therefore we can look for
symmetric and antisymmetric solutions. However, the Hamiltonian does not
commute with the exchange operator $\Omega \hat{P}$. With the parity as a good quantum number
it is straightforward to generalize the theorem of Section \ref{boundgeneral} to include exchange. To see
this, take $\rho_M\equiv \max(\rho,\rho_{\mathrm{ex}})$. If $u(z)$ is
symmetric, the exchange 
shifts the potential to $V(|z|)+\Omega(|z|)$, while if $u(z)$ is antisymmetric
it shifts the potential to $V(|z|)-\Omega(|z|)$. Therefore, obtaining the
bound states of $H_{\mathrm{ex}}$ reduces again to a polynomial equation of degree
$2\rho_M\pm 1$, and all the results of our theorem apply by changing $\rho$ by
$\rho_M$ and $V$ by $V\pm \Omega$. However, it is no longer true that a
hardcore condition $|V(0)|\to \infty$ maps ``bosons''
onto ``fermions'' (symmetric onto antisymmetric solutions). Indeed, the non-trivial dependence of $H_{\mathrm{ex}}$ on the
parity of the eigenstates makes it possible to have states above as well as
below its continuum even if $V$ and $\Omega$ have both the same definite
sign. This is the fact that makes $H_{\mathrm{ex}}$ violate the hypotheses of
the BFMT \cite{Girardeau}, and
it explains the appearance of exotic three-body bound states on a 1D lattice
\cite{MVDPAS}.

%%%%%%%%%%%%%%%%%%%%%%%%%%%%%%%CONCLUSIONS%%%%%%%%%%%%%%%%%%%%%%%%%%%%%%%
\section{Conclusions}

In this paper, we have shown how the exact wave functions and energies
of any bound state
of two particles on a one-dimensional tight-binding lattice can be calculated
by solving a polynomial equation whose order increases slowly with increasing
range of the two-body interaction potential, which can also include
parity-dependent terms, such as effective particle exchange. We have
shown that the calculation of the exact scattering states is possible, and simple. We have also shown how the
zero-energy resonances associated with the entry or exit of a bound state can
be trivially and exactly located, and related the scattering lengths to the
number of bound states above or below the continuum by making reference to the
discrete version of Sturm theory.

There are, on the other hand, many open problems in the physics of few particles
on a lattice. At the two-body level, there is still no general prescription for the
calculation of bound and scattering states on two- and three-dimensional lattices with
arbitrary finite range interactions. For the
three-body case, the Efimov effect \cite{Efimov}, which was shown to
appear on a three-dimensional (3D) simple cubic lattice \cite{RevMattis}, has not
been quantitatively examined yet; the lattice Efimov effect should depend largely
on the total quasi-momentum of the system and, moreover, there
should appear new kinds of exotic three-body bound states in 3D with no analog
in continuous space, as has been shown to be the case in 1D
\cite{MVDPAS}. It would also be very interesting to explore few-body effects
in other lattice geometries.

%%%%%%%%%%%%%%%%%%%%%THANKS%%%%%%%%%%%%%%%%%%%%%%
\begin{acknowledgments} 
Useful discussions with David Petrosyan, Luis Rico and especially Gerald Teschl
are gratefully acknowledged. 
This work was supported by the EU network EMALI.
\end{acknowledgments} 

%%%%%%%%%%%%%%%%%%%%%%%%%%%%%EQUIVALENCE%%%%%%%%%%%%%%%%%%%%%%%%%%%%%%%%%%
\appendix

\section{Equivalence between repulsive and attractive potentials}\label{app}

We state a general result for an $N$-body system on a hypercubic lattice in any dimension
which, although being usually implicitly assumed, is useful and important to keep in mind,
specially when dealing with purely attractive or repulsive two-body
interactions.  

Let $H$ be the following second-quantized Hamiltonian
\be
H=-J\sum_{\langle\mathbf{m},\mathbf{n}\rangle}\hat{a}^{\dagger}_{\mathbf{m}}
\hat{a}_{\mathbf{n}} + \hat{F}(\{ \hat{N}\}),\label{HHam}
\ee 
where $J$ is the single-particle tunneling rate, $\langle\mathbf{m},\mathbf{n}\rangle$
denotes that the sum runs only through nearest neighbors,
$\hat{a}^{\dagger}_{\mathbf{n}}$ ($\hat{a}_{\mathbf{n}}$) is the creation
(annihilation) operator of a single particle (boson or fermion) at site
$\mathbf{n}$ and where $\hat{F}$
is an arbitrary analytic function of the number operators at each site
$\mathbf{n}$, $\hat{N}_{\mathbf{n}}\equiv
\hat{a}^{\dagger}_{\mathbf{n}}\hat{a}_{\mathbf{n}}$.\\
If $\mathbf{n}\equiv (n_1,n_2,\ldots,n_d)$ is a point of a
$d$-dimensional hypercubic lattice, we define the unitary operation $\hat{G}$
so that $\hat{G}=\hat{G}^{\dagger}=\hat{G}^{-1}$, with the actions 
\begin{equation}
\hat{G}\hat{a}^{\dagger}_{\mathbf{n}}=(-1)^{\sum_{s=1}^{d}
  n_s}\hat{a}^{\dagger}_{\mathbf{n}}.\label{trans}
\end{equation}
We easily see that $H-2\hat{F}=-\hat{G} H \hat{G}^{-1}$ is unitarily equivalent to
$-H$. This 
implies that the spectrum of a Hamiltonian containing the potentials
included in $\hat{F}$ is obtained by changing the sign of every point
in the spectrum of the corresponding Hamiltonian having $\hat{F}$ replaced by $-\hat{F}$. In the case
of $\hat{F}$ containing purely repulsive (or attractive) two-body interactions,
this result implies the formal equivalence between attractive and repulsive
potentials.

%%%%%%%%%BIBLIOGRAPHY%%%%%%%%%%%%%


\begin{thebibliography}{99}

\bibitem{Ashcroft}  
N.J.~Ashcroft and and N.D.~Mermin,
{\it Solid State Physics} 
(International Thomson Publishing, New York, 1976).

\bibitem{Hubbard}
J. Hubbard, 
Proc. Roy. Soc. A {\bf 276} 238 (1963).

\bibitem{bosMI} 
M.P.A. Fisher, P.B. Weichman, G. Grinstein, and D.S.~Fisher, 
Phys. Rev. B {\bf 40}, 546 (1989).

\bibitem{Jaksch} 
D. Jaksch {\it et al.},
Phys. Rev. Lett. {\bf 81}, 3108 (1998).

\bibitem{OptLatRev}
O.~Morsch and M.~Oberthaler,
Rev. Mod. Phys. {\bf 78}, 179 (2006);
I.~Bloch, J.~Dalibard, and W.~Zwerger,
Rev. Mod. Phys. {\bf 80}, 885 (2008);
M.~Lewenstein {\it et al.}, 
Adv. Phys. {\bf 56},  243 (2007).

\bibitem{transition}
M. Greiner {\it et. al.},
Nature {\bf 415}, 39 (2002).

\bibitem{Winkler}
K. Winkler {\it et al.},
Nature {\bf 441}, 853 (2006).

\bibitem{PiilMolmer}
R. Piil and K. M\o lmer,
Phys. Rev. A {\bf 76}, 023607 (2007).

\bibitem{MVDP1}
M. Valiente and D. Petrosyan,
J. Phys. B {\bf 41}, 161002 (2008). 

\bibitem{MVDP2}
M. Valiente and D. Petrosyan,
J. Phys. B {\bf 42}, 121001 (2009).  

\bibitem{MVDPAS}
M. Valiente, D. Petrosyan and A. Saenz,
Phys. Rev. A {\bf 81}, 011601(R) (2010).   

\bibitem{weiss}
C. Weiss and H.-P. Breuer,
Phys. Rev. A {\bf 79}, 023608 (2009).

\bibitem{JinSong}
L. Jin, B. Chen and Z. Song,
Phys. Rev. A {\bf 79}, 032108 (2009).

\bibitem{PiilMolmerdifferentmass}
R. T. Piil, N. Nygaard, and K. M\o lmer,
Phys. Rev. A {\bf 78}, 033611 (2008).

\bibitem{RevMattis}
D. C. Mattis,
Rev. Mod. Phys. {\bf 58}, 2, 361 (1986).

\bibitem{footnote1}
In continuous space this is equivalent to particles with different masses.

\bibitem{Martikainen}
J.-P. Martikainen,
Phys. Rev. A {\bf 78}, 035602 (2008).

\bibitem{Girardeau}
M.D. Girardeau,
J. Math. Phys. {\bf 1}, 516 (1960).

\bibitem{PSAF}
D. Petrosyan {\it et al.},
Phys. Rev. A {\bf 76}, 033606 (2007).

\bibitem{Damanik}
D. Damanik, D. Hundertmark, R. Killip and B. Simon,
Commun. Math. Phys. {\bf 238}, 545 (2003).

\bibitem{Teschl}
G. Teschl, 
{\it Jacobi operators and completely integrable nonlinear lattices}
(Mathematical surveys and monographs, American Mathematical Society, 2000).

\bibitem{Trefzger1}
C. Menotti, C. Trefzger and M. Lewenstein,
Phys. Rev. Lett. {\bf 98}, 235301 (2007).

\bibitem{Trefzger2}
C. Trefzger, C. Menotti and M. Lewenstein,
Phys. Rev. A {\bf 78}, 043604 (2008).

\bibitem{Menotti}
Th. Lahaye, C. Menotti, L. Santos, M. Lewenstein and T. Pfau,
Rep. Prog. Phys. {\bf 72}, 126401 (2009).

\bibitem{footnote2}
Note that since $V(z)\le 0$ for all $z\in\mathbb{Z}$, there can be no roots
for $-1<\alpha<0$.

\bibitem{HundertmarkSimon}
D. Hundertmark and B. Simon,
J. Approx. Theory {\bf 118}, 106 (2002).

\bibitem{BetheI}
M. Karbach and G. M\"uller,
Comp. in Phys. {\bf 11}, 36 (1997).

\bibitem{PiilMolmerFeshbach}
N. Nygaard, R. Piil and K. M{\o}lmer,
Phys. Rev. A {\bf 78}, 023617 (2008).

\bibitem{footnote3}
Note that we have normalized the Hamiltonian as $\tilde{H}
= H/|J^{(K)}|$.

\bibitem{footnote4}
If the calculation is implemented with a symbolic package,
  the value of $V(0)$ at which the resonance occurs can be calculated exactly
  (with no machine precission limit). In our example, $V(0)$ is a rational
  number whose (large) numerator and denominator can be calculated exactly
  this way.

\bibitem{footnote5}
On the lattice, a function $f$ is said to have a node between $n$
  and $n+1$ iff $f(n)f(n+1)<0$.

\bibitem{Efimov}
V.N. Efimov,
Phys. Lett. B {\bf 33}, 563 (1970).













\end{thebibliography}
\end{document}